\documentclass[twocolumn,prl]{revtex4-2}
\usepackage{hyperref}
\usepackage[hyphenbreaks]{breakurl}
\hypersetup{breaklinks=true,colorlinks=true, citecolor=blue, urlcolor=blue, linkcolor=blue}
\PassOptionsToPackage{hyphens}{url}
\usepackage{xr}  % 如果使用了 hyperref 宏包，请改用 \usepackage{xr-hyper}
\usepackage{CJK}
\usepackage{mathrsfs}
\usepackage{amssymb}
\usepackage{amsmath}
\usepackage{latexsym}
\usepackage{graphicx}
\usepackage{dcolumn}
\usepackage{bm}
\usepackage{graphicx}
\usepackage{float}
\usepackage[normalem]{ulem}
\usepackage{color,xcolor}
\usepackage{multirow}
\usepackage{enumerate}
\usepackage{ulem}
\usepackage{textcomp}

\newcolumntype{d}[1]{Dc{.}{.}{#1}}
\begin{document}
	\begin{CJK*}{UTF8}{}
		
		\title{ 
			General Radial-Composition Correlations in Two-Component Many-Body Systems
		}
		\author{Y. Lei ({\CJKfamily{gbsn}雷杨})}
		\email[]{leiyang1985@swust.edu.cn}
		\affiliation{School of Nuclear Science and Technology, Southwest University of Science and Technology, Mianyang 621010, China}
		\date{\today}
		
		\begin{abstract}
			The linear correlation between RMS radius difference and composition asymmetry in two-component many-body systems is a robust feature observed across nuclear experiments, diverse nuclear structural models, molecular dynamic simulations for bimetallic clusters, and galactic modeling with self-interacting dark matter. We identify the short-range attractive central force as the key ingredient for its emergence, a mechanism underpinned by the coordinate transformation under low-energy harmonic-oscillator approximation, the virial theorem, and Pauli principle/hard core potential, in many-fermion system/classic many-body system.
		\end{abstract}
		%\pacs{XXX}
		\maketitle
	\end{CJK*}

	%\section{Introduction}\label{sec-int}	
	Complex bound many-body systems, such as nuclei, metallic nanoclusters, and galaxies, are central to understanding the universe and enabling practical applications \cite{Sherrill01042005}. Remarkably, although their interactions are complex and often not fully known, such systems exhibit striking regularities from a reductionist perspective. Examples include the nuclear $N_pN_n$ systematics \cite{R_F_Casten_1996}, Wigner-Dyson level statistics in metallic clusters and quantum dots \cite{PhysRevB.70.205411,PhysRevB.75.205107,PhysRevLett.77.1123,PhysRevLett.80.4522}, and scaling relations for galaxy clusters and groups \cite{10.1093/mnras/222.2.323,Giodini2013}. The simplicity emerging from complexity often signals underlying dynamical and symmetry critical points, which merits deeper understanding.
	
	For nuclear many-body system, random-interaction ensembles provide a unique and powerful framework for identifying robust patterns in nuclear systems \cite{KOTA2001223,ZHAO20041,ZELEVINSKY2004311,RevModPhys.81.539}. By performing many-body calculations for pesudo nuclei with randomized interaction matrix elements, one can generate ensembles of resultant observables. Thus, features that emerge with statistical prevalence in such ensembles are considered independent of specific interaction details, such as the spin-zero and positive-parity ground states \cite{PhysRevLett.80.2749,PhysRevC.70.054322}, collective-like motions \cite{PhysRevLett.84.420,PhysRevC.83.024302,PhysRevC.90.064313,PhysRevC.104.054319,PhysRevResearch.5.013109,4qrd-1dqw}, and odd-even staggering in proton-neutron interactions \cite{PhysRevC.91.054319}. Furthermore, specific sampling in random-interaction ensembles may reveal the key interaction features driving robust patterns, as Refs.  \cite{PhysRevC.83.044302,PhysRevC.93.024319,Qin2018}.

	%above is chatted****************************************************
	
	Using this random-interaction approach\footnote{Calculation details are all described in the 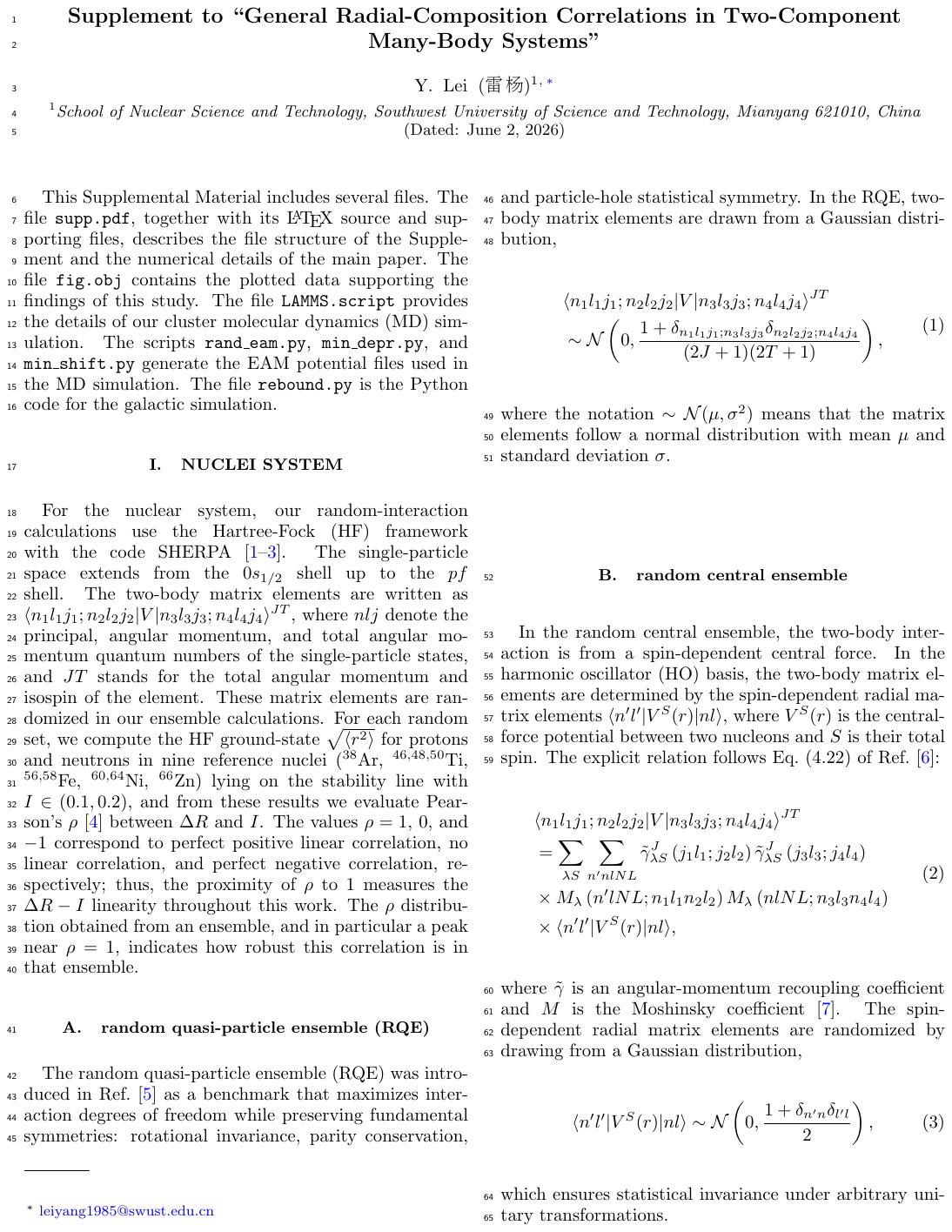 of the Supplement Material \protect\cite{supp}.}, we study a general correlation between matter distribution and composition in two-component systems, from nuclei of protons and neutrons to galaxies of luminous and dark matter. We show that the difference between the root-mean-square (RMS) radii of the two components, $\Delta R=\sqrt{\langle r^2\rangle_1}-\sqrt{\langle r^2\rangle_2}$, is linearly correlated with the composition asymmetry, $I=(p_1-p_2)/(p_1+p_2)$, where $\sqrt{\langle r^2\rangle_i}$ and $p_i$ are the RMS radius and fraction of component $i$. Although this correlation has been observed in nuclear systems~\cite{PhysRevLett.87.082501,rch-I,MYERS1969395,MYERS1974186,MYERS1980267,PETHICK1996173,qvsg-hqxm}, it has remained unexplored in other systems. Here, we identify its key mechanism and employ various many-body methods to verify it at different scales. This scale-independent correlation may benefit future simulations, experiments, and observations of diverse many-body systems.
	
	%chatted****************************************************

	%\section{Nuclear systems}
	%\subsection{$\Delta R$: neutron skin thickness}
	In nuclear physics, the quantity $\Delta R$ defined above is known as the neutron skin thickness. It serves as the only laboratory proxy on earth, for neutron-matter stiffness and neutron-star radius \cite{brown2000,PhysRevLett.86.5647,PhysRevC.86.015803,Hagen2016,brown-rch,PhysRevLett.120.172702,PhysRevC.100.015802}. Specifically, $\Delta R$ is found to be positively correlated with the symmetry energy slope $L$ \cite{rnp-1,rnp-2,explain_phys_rep}, a parameter that plays a decisive role in characterizing neutron star radii.
	
	%chatted*********************
	
	\begin{figure}[!htb]
		\includegraphics[angle=0,width=0.45\textwidth]{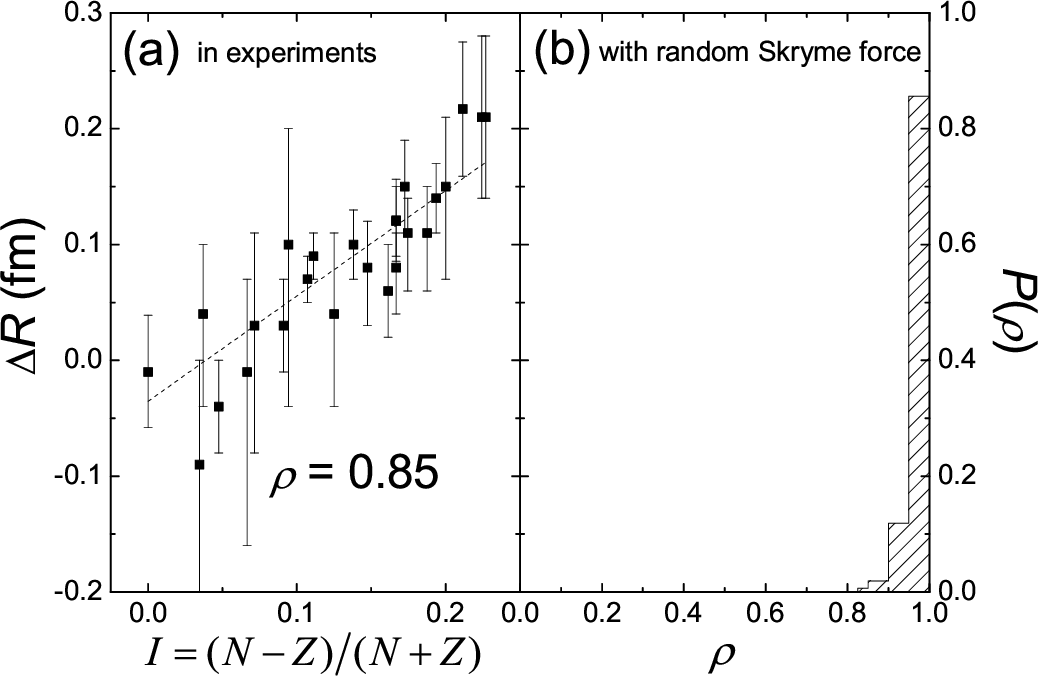}
		\caption{
			Nuclear $\Delta R-I$ correlation. (a) Experimental data \cite{zenihiro2018directdeterminationneutronskin,doi:10.1142/S0218301304002168,PhysRevC.46.1825,PhysRevLett.129.042501,PhysRevLett.131.202302} give Pearson's $\rho=0.85$ \cite{pearson1895}. (b) $\rho$ distribution from random Skyrme-Hartree-Fock calculations \cite{qvsg-hqxm}; the peak near $\rho\simeq0.9$ shows robustness under random interactions.
			%chatted*********************
		}\label{fig:exp}
	\end{figure}
	
	Similarly, the quantity $I$ defined above corresponds to the nuclear isospin asymmetry $(N-Z)/(N+Z)$. It has long been known that $\Delta R$ is positively correlated with $I$, as shown by experiments in Fig. \ref{fig:exp}(a). This linear trend is also robustly reproduced across diverse theoretical frameworks, e.g., liquid drop models, mean-field theories, and $ab~initio$ approaches \cite{PhysRevLett.87.082501,rch-I,MYERS1969395,MYERS1974186,MYERS1980267,PETHICK1996173}. We quantify this correlation with Pearson's $\rho$ \cite{pearson1895}: $\rho=1$ (0) for perfect (no) linearity. The experimental data yield $\rho=0.85$, as shown in Fig. \ref{fig:exp}(a), confirming a robust linear trend. Remarkably, we recently noted that this correlation persists even under random parametrizations of the Skyrme force \cite{qvsg-hqxm}, as shown in Fig. \ref{fig:exp}(b). The robustness of this correlation across various theoretical calculations already suggests that this linearity may be largely independent of specific modeling details, and instead arises from few but intrinsic properties of the nuclear system.
	
	%above is chatted****************************************************
	
	%\subsection{correlation emerging out of random interactions}
	
	\begin{figure}[!htb]
		\includegraphics[angle=0,width=0.45\textwidth]{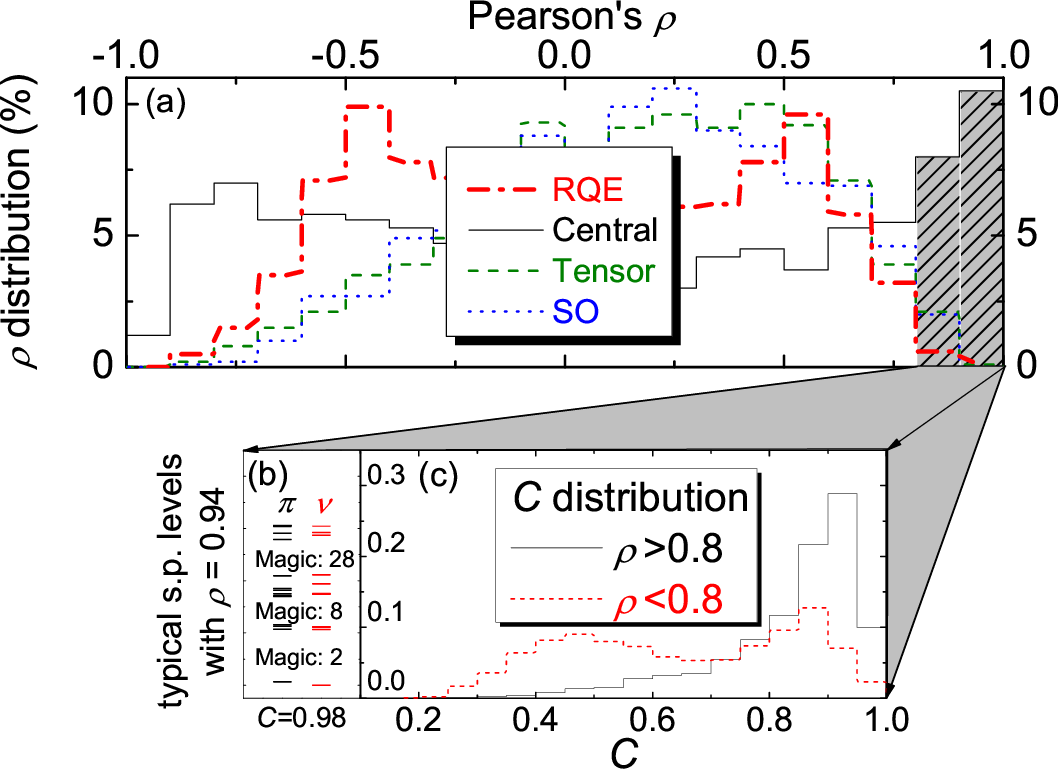}
		\caption{(Color online)
			Nuclear $\Delta R-I$ correlation in random ensembles. (a) $\rho$ Distribution for the $\Delta R-I$ correlation across different random ensembles: RQE, random central, tensor, and spin-orbit (SO) ensembles. The shaded region highlights the peak near $\rho=0.9$ for the random central ensemble, from which data for (b) and (c) are taken. (b) Proton and neutron level schemes for a pseudo nucleus, $^{38}$Ar, in a representative sample ($\rho = 0.94$), revealing clear gaps at harmonic oscillator (HO) magic numbers (2, 8, 28). The HO-likeness measure $C$ for this pseudo $^{38}$Ar, defined as the product of Pearson coefficients for proton and neutron spectra with exact 3D HO level scheme, equals to 0.98, suggesting a strong similarity of the spectra under investigation to HO ones. (c) $C$ distributions for pseudo nuclei with $\rho>0.8$ and $\rho<0.8$ in the random central ensemble. Samples with $\rho>0.8$ cluster at $C\simeq 0.9$, suggesting HO-like nature.
			%chatted***********************
		}\label{fig:r-dis}
	\end{figure} 
	
	To test the robustness of the $\Delta R-I$ correlation, we perform Hartree-Fock (HF) calculations \cite{PhysRevC.66.034301,PhysRevC.67.044315,PhysRevC.69.024311} with several random ensembles. Calculation detail is described in Sec. \ref{supp-sec:nucl}. We start with the random quasi-particle ensemble (RQE) \cite{PhysRevLett.80.2749}, where two-body matrix elements are drawn randomly according to Eq.~(\ref{supp-eq:rqe}) of supp.pdf in the Supplemental Material \cite{supp}, to maximize interaction freedom while preserving nuclear symmetries. For each RQE sample, Pearson's $\rho$ is calculated with resultant $\Delta R$ and $I$ of selected reference nuclei. Obtained $\rho$ distribution appears in Fig. \ref{fig:r-dis}. The RQE yields a negligible probability (less than 1\%) of strong correlation with $\rho>0.9$, indicating that the $\Delta R-I$ linearity may not be guaranteed by nuclear symmetries but require specific physical drives in nucleon-nucleon interactions.
	
	%above is chatted****************************************************
	We test whether central, tensor, or spin-orbit forces drive the $\Delta R-I$ correlation, separately. These interaction types encompass all interactions that are independent of or linearly dependent on momentum, and compatible with general nuclear symmetries, specified by radial matrix elements of radial potential between nucleons \cite{lawson1980theory}. We randomize these elements according to Eqs. (\ref{supp-eq:central}), (\ref{supp-eq:tensor}), and (\ref{supp-eq:so}) of supp.pdf in the Supplemental Material \cite{supp}, creating random central, tensor, and spin-orbit ensembles, respectively. The resulting $\rho$ distributions are shown in Fig.~\ref{fig:r-dis}. The random tensor and spin-orbit ensembles, like the RQE, fail to produce significant correlations with $P(\rho>0.9) \approx 0$. In stark contrast, the random central ensemble exhibits a pronounced peak at $\rho \approx 0.95$, with nearly 20\% of samples displaying strong linearity ($\rho>0.8$). This statistical filtration unambiguously identifies the central force as the primary driver of the $\Delta R-I$ correlation.
	
	%above is chatted****************************************************
	
	%\subsection{harmonic oscillator feature from sampling}
	
	Despite the central force being the primary driver, not all samples in the random central ensemble exhibit a strong $\Delta R-I$ correlation; additional features must distinguish those that do. To identify them, we select about 2000 samples with $\rho>0.8$ from the 10000 in the ensemble and examine their HF single-particle spectra. Figure \ref{fig:r-dis}(b) shows a typical example for $^{38}$Ar with $\rho=0.94$, displaying clear shell closures at the harmonic-oscillator (HO) magic numbers (2, 8, 28) and nearly degenerate levels in the lowest two shells, resembling an HO spectrum. We quantify this HO-likeness by $C$, the product of Pearson coefficients between the proton (neutron) spectrum and the exact 3D HO spectrum; $C=0.98$ for the example in Fig. \ref{fig:r-dis}(b). We then compute $C$ for all samples with $\rho>0.8$ and $\rho<0.8$ shown in Fig. \ref{fig:r-dis}(c). The $\rho>0.8$ samples, which roughly reproduce the $\Delta R-I$ correlation, are significantly more HO-like, with $C$ peaked near 0.9, and vice versa. This suggests that an HO-like mean field underlies the nuclear $\Delta R-I$ correlation.
	
	%chatted***********************
	
	To directly examine the influence of the HO potential, we use a parameterized central-force HO potential $V^S(r)= V_0(r-r_0)^2$ for $S=0,~1$, and apply it to the same reference nuclei as in the random central ensemble. We find that a repulsive HO force ($V_0 < 0$) consistently yields a roughly negative correlation ($\rho \approx -0.8$), contrary to nuclear observations. For attractive forces ($V_0 > 0$), the linearity is remarkably robust against the potential sharpness ($V_0$), yet sensitive to the potential minimum $r_0$. Strong positive correlations ($\rho \approx 1$) only emerge when the minimum is localized within $r_0 < 1.5\sqrt{\hbar/m\omega}$; as $r_0$ increases, the correlation rapidly degrades to $\rho \approx -0.8$. We call a minimum near zero ``short range". Our results then show that the $\Delta R-I$ linearity is a hallmark of an attractive, short-range central HO interaction.
	
	%chatted***********************
	
	%\subsection{explanation based on harmonic oscillator}
	
	The emergence of the nuclear $\Delta R-I$ correlation from an attractive short-range HO potential may be supported by three pillars: the center-of-mass transformation, the virial theorem, and the Pauli principle. The center-of-mass transformation maps an HO interacting $N$-nucleon system to independent particles in a global HO mean field \cite{10.1098/rspa.1955.0239}. According to the virial theorem, in the HO mean field, the energy and potential, i.e., mean square radius, of HF single-particle states are proportional to each other. With the Pauli principle, nucleons sequentially occupy levels with increasing energy and, consequently, increasing spatial extent. The accumulation of like nucleons thus drives a systematic expansion of the total RMS radius. Thus, the asymmetry between proton and neutron numbers ($I$) is inherently correlated to their RMS-radius difference ($\Delta R$).
	
	%chatted**************************************
	
	Realistic nuclear forces, though more complex than a zero-range HO, still induces the $\Delta R-I$ correlation \cite{MYERS1969395,MYERS1974186,MYERS1980267,PETHICK1996173}. That may be because this correlation come from nuclear ground states. In ground states, the low-energy minimum of any short-range potential can be approximated to a HO potential centered near $r=0$, leading to the $\Delta R-I$ correlation. For example, random Skyrme forces \cite{qvsg-hqxm}, with dominant attractive zero-range central forces \cite{10.1098/rspa.1961.0018,SKYRME1958615}, is expected to frequently yield this correlation as shown in Fig.~\ref{fig:exp}(b). Thus, the key ingredient for the $\Delta R-I$ correlation in ground states of many-fermion system could be a short-range attractive central force.
	
	%chatted**********************************
	
	%\subsection{verification for short-range attraction as the key ingredient for nuclear $\Delta R-I$ correlation}
	
	\begin{figure}[!htb]
		\includegraphics[angle=0,width=0.45\textwidth]{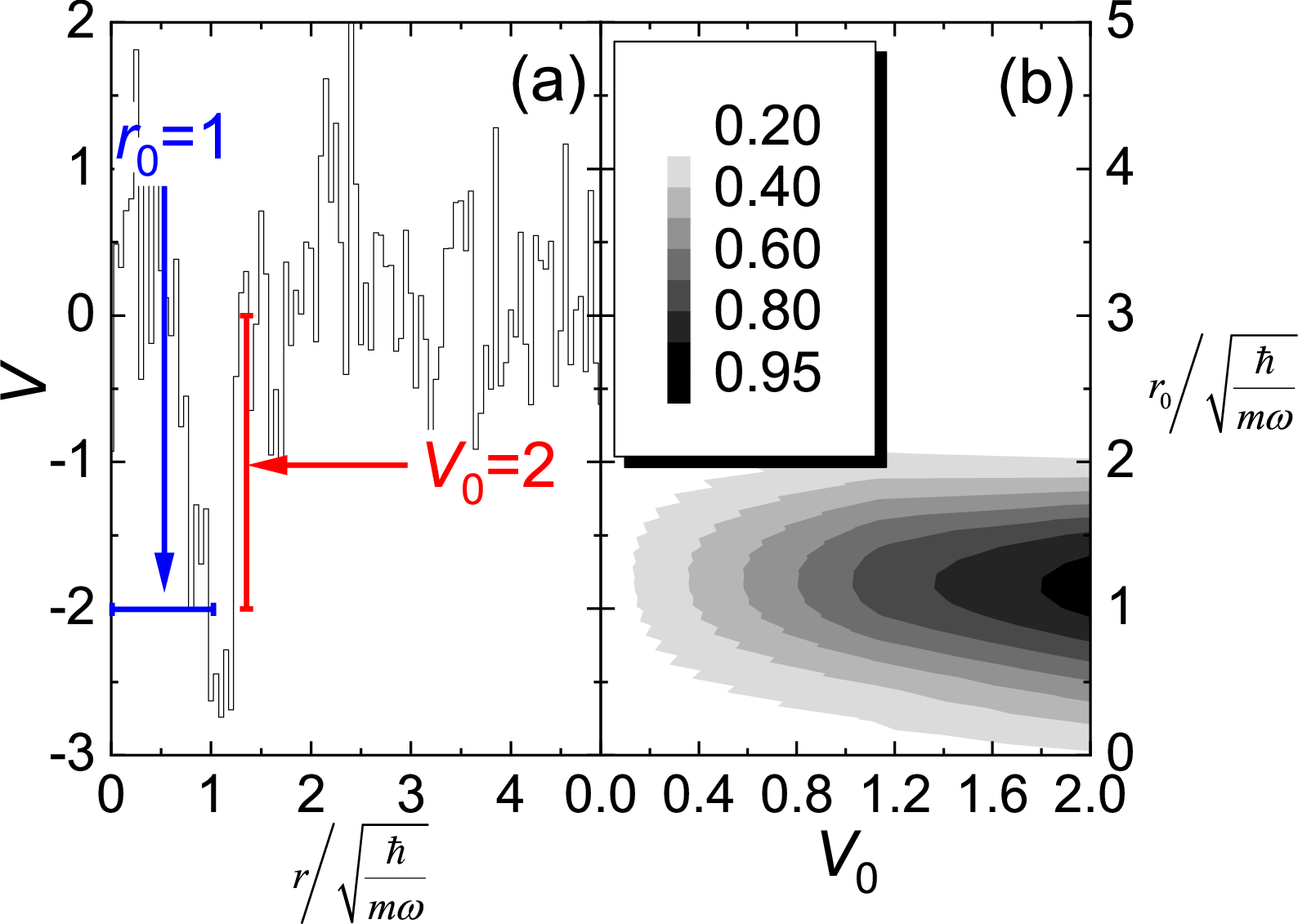}
		\caption{
			(Color online) Random short-range ensemble. (a) Example potential ($S=0$) at $V_0=2$, $r_0=1$, with $V_0$ ($r_0$) controlling degree of minimum attraction (minimum range). (b) $P(\rho>0.9)$ as a function of $V_0$ and $r_0$. The peak near $(V_0=2,r_0=1)$ indicates that a short-range, attractive central force drives the $\Delta R-I$ correlation in nuclear ground states.
		%chatted*************
		}\label{fig:rho-v0-r0}
	\end{figure}
	
	To test whether a short-range attractive central force produces the $\Delta R-I$ correlation in many-fermion systems, we construct ``random short-range ensemble" with random spin-independent radial potentials, parameterized by degree of minimum attraction $V_0$ and minimum range $r_0$ according to  Eq. (\ref{supp-eq:v_short}) of supp.pdf in the Supplemental Material \cite{supp}. Fig.~\ref{fig:rho-v0-r0}(a) shows a typical radial potential in such an ensemble. Random HF calculations yield $P(\rho>0.9)$, the probability of $\rho>0.9$, plotted against $V_0$ and $r_0$ in Fig.~\ref{fig:rho-v0-r0}(b). The single peak at $(V_0=2,r_0=1)$ confirms that a short-range attractive central force is essential for the $\Delta R-I$ correlation.
	
	%chatted*****************************************
	%\section{classic systems}
	Although our discussion has been limited to nuclear quantum systems, the three pillars (coordinate transformation, virial theorem, and Pauli principle) supporting the explanation for the $\Delta R-I$ correlation may also apply classically. The first two pillars are valid for both quantum and classic systems. For the third one, in classical models of many-fermion systems, the Pauli principle can be mimicked by a hard core near $r=0$. A hard core forbids spatial overlap and converts particle number into spatial extent, much like the Pauli principle. Hence, a classical many-body system with short-range attraction and a hard core may also exhibit the $\Delta R-I$ correlation in low-energy states.
	
	%chatted*********************
	
	%\subsection{bimetallic clusters}
	
	Molecular dynamics (MD) \cite{B703897F} simulation under Newton's law for Cu$_x$Al$_y$ nanoalloy clusters supports classic $\Delta R-I$ correlation and its mechanism. Taking such clusters as many-body systems, $\Delta R$ is the RMS radius difference between Cu and Al, and $I=(x-y)/(x+y)$ the composition asymmetry by atom count. We perform MD simulations with LAMMPS \cite{LAMMPS} to obtain the time-ergodic ground-state $\Delta R$ for nine reference CuAl clusters with different $I$, giving a Pearson's $\rho$ for each EAM potential. Calculation detail are described in Sec. \ref{supp-sec:md} of supp.pdf in the Supplemental Material \cite{supp}.
	
	%chatted*********************
	
\begin{figure}[!htb]
		\includegraphics[angle=0,width=0.45\textwidth]{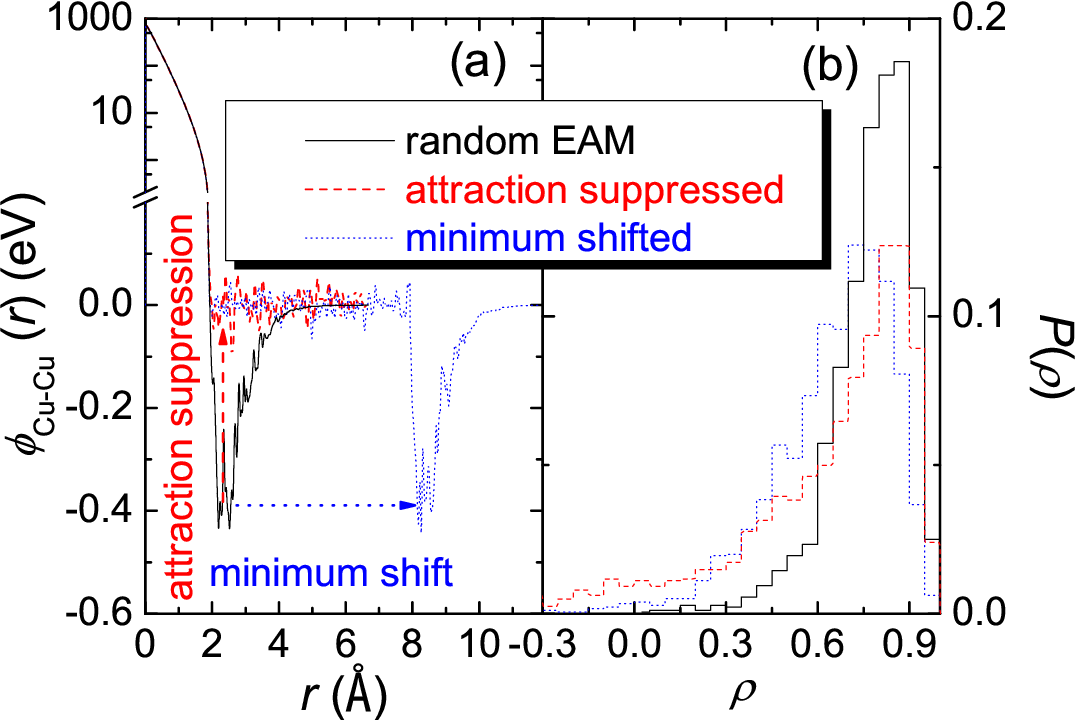}
		\caption{(Color online)
			MD simulated CuAl clusters within random ensembles.	(a) Typical Cu-Cu pair potentials from the random EAM, attraction-suppressed, and minimum-shifted ensembles. (b) Pearson's $\rho$ distributions for the three ensembles.
			%chatted***********************************
		}\label{fig:cual}
	\end{figure}

	Alloy MD simulation usually adopts embedded-atom method (EAM) potential \cite{10.1115/1.3183784}, where the pair potential carries most of the central force between atoms. Following random interaction approach, we randomize the pair potential to create simulation ensembles, to test the robustness of the $\Delta R-I$ correlation in a classical many-body system and to probe the role of short-range attraction. Based on the CuAl EAM potential \cite{PhysRevB.54.8398} from the repository \cite{Hale_2018,BECKER2013277,nist_interatomic_potentials}, we add Gaussian fluctuations to the attractive part as described by Eq.~(\ref{supp-eq:rand-eam}) of supp.pdf in the Supplemental Material \cite{supp} to generate a ``random EAM'' ensemble. A typical pair potential from this ensemble is shown in Fig.~\ref{fig:cual}(a); it remains short-ranged and attractive.
	%chatted*********************
	
	Random EAM ensemble frequently yields strong correlations with $P(\rho>0.85)\approx 50\%$, and a peak near $\rho\approx0.9$, as demonstrated by the resulting $\rho$ distribution plotted in Fig.~\ref{fig:cual}(b).	To examine the correlation's origin, we modify the pair potentials from Ref.~\cite{PhysRevB.54.8398} in two ways before randomization, yielding two ensembles. In the ``attraction-suppressed'' ensemble the attractive part is set to zero. In the ``minimum-shifted'' ensemble the attractive part is shifted outward by 6\,\AA\ to create a long-range attraction, with the gap filled by zero. Both ensembles are then randomized as for random EAM ensemble. Figure~\ref{fig:cual}(a) shows representative Cu-Cu potentials, and Fig.~\ref{fig:cual}(b) gives their $\rho$ distributions. Suppressing the attraction or making it long-range reduces $P(\rho>0.85)$ by nearly 40\% and shifts the peak of the distribution within minimum-shifted ensemble. These results confirm that, even in a classical many-body system, a short-range attractive central force is essential for the $\Delta R-I$ correlation.
	
	%chatted********************************	 
	
	%\subsection{galactic systems}
	
	The emergency mechanism of the $\Delta R-I$ correlation from a short-range attractive central force of classic many-body systems are supported by three pillars: coordinate transformations, the virial theorem, and a hard core between particles. To test this mechanism, one can remove one pillar and see if the correlation persists. Since the first two are generic, we focus on removing the hard core. This is difficult because a hard core ensures incompressibility of many-body system and prevents collapse into a singularity. For instance, in MD simulations of AlCu clusters, the EAM potential inherently supplies a strong repulsive core, and suppressing it will lead to numerical instabilities in LAMMPS~\cite{BEUTLER1994529}. Fortunately, in a two-component system one can remove the hard core for only one species, relying on the other to maintain global incompressibility. In a classical system, coreless particles with zero non-gravitational cross section must have nonzero mass to move under Newton's laws, thus resembling weakly interacting massive particles, a dark matter candidate \cite{Roszkowski_2018}. Then, we model the remaining component as ``ordinary'' matter undergoing elastic hard-ball collisions, characterized by a force range $r_o=0.2$, while the dark matter particles interact only gravitationally, i.e., $r_d=0$. Then gravity is introduced as a short-range central attraction acting on both species, constructing a minimal galactic model. Using the $N$-body integrator REBOUND~\cite{refId0,10.1093/mnras/stu2164}, we simulate this self-gravitating system, neglecting relativity, and compute Pearson's $\rho$ between $\Delta R$, the difference between RMS radii of ordinary matter and dark matter, and the real-time composition asymmetry $I$ for different initial $I$ (details in Sec.~\ref{supp-sec:gal} of supp.pdf in the Supplemental Material~\cite{supp}). As expected, for $r_d=0$ $\rho$ drifts away from 1 after initial fluctuation.
	
	%chatted********************************	 
	
	\begin{figure}[!htb]
		\includegraphics[angle=0,width=0.45\textwidth]{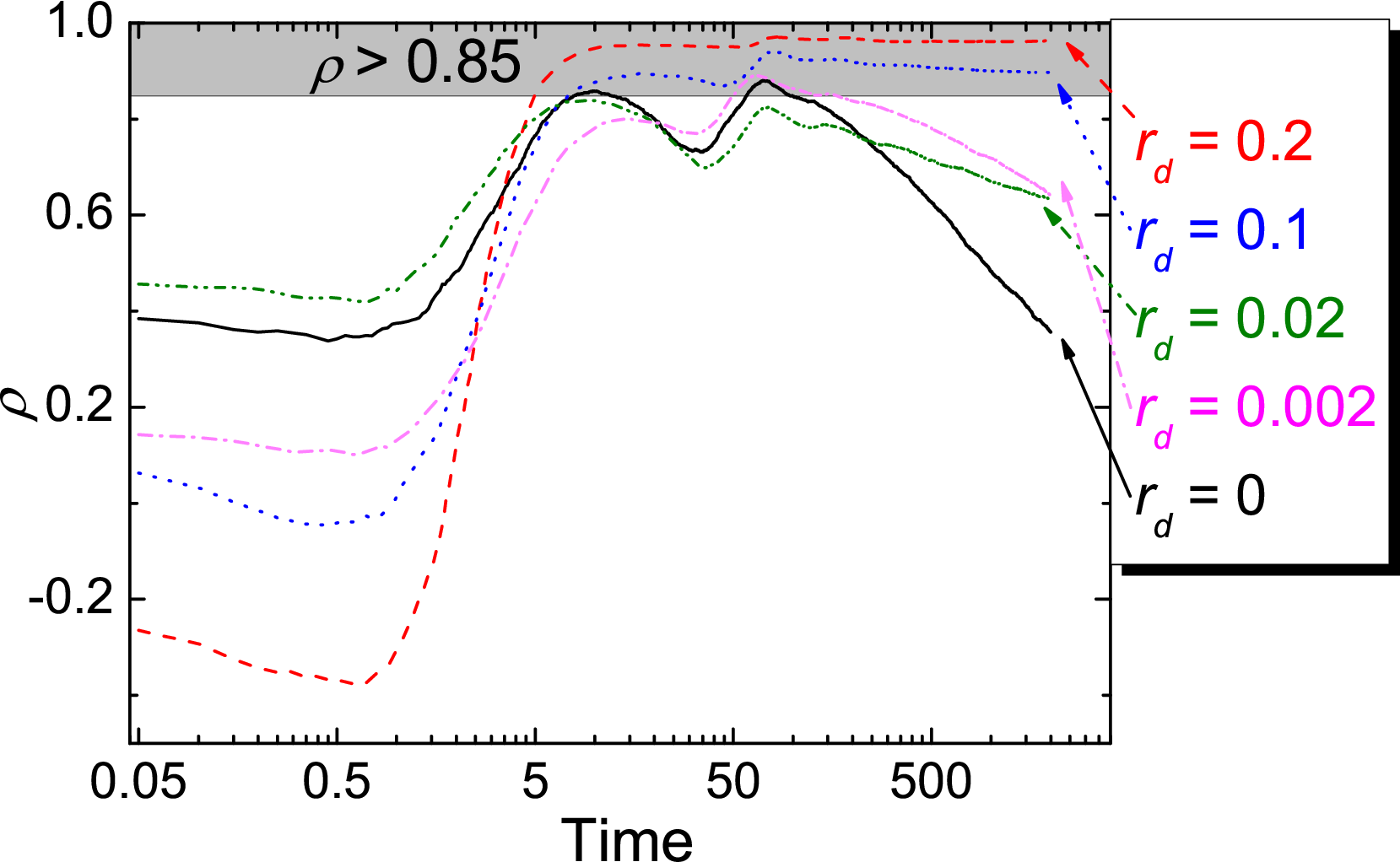}
		\caption{(Color online)
			Time evolution of Pearson's $\rho$ for galaxy models with different initial composition asymmetry $I$ and dark-matter hard-ball force range $r_d$. Time is shown on a log scale. The ordinary particles have $r_o=0.2$. Five $r_d$ values are considered: 0, 0.002, 0.02, 0.1, and 0.2.
			%chatted
		}\label{fig:dm}
	\end{figure}
	
	Figure~\ref{fig:dm} shows the time evolution of $\rho$ for reference galaxies. For $r_d=0$, $\rho$ oscillates strongly before $t\sim100$ and then decreases steadily, deviating from linearity after $t\sim500$, as expected. Having shown that removing the hard core destroys the correlation, we now verify that restoring it recovers $\Delta R-I$. We assign a nonzero hard-ball force range $r_d$ to dark matter, corresponding to strongly self-interacting dark matter recently proposed to address small-scale structure problems \cite{PhysRevLett.113.171301,PhysRevLett.125.131301,PhysRevLett.128.172001}. Dark matter and ordinary matter still interact only via gravity. When $r_d$ is comparable to $r_o$ ($r_d=0.1$ or 0.2), the evolution stabilizes with $\rho>0.85$, indicating that the correlation reappears. For $r_d=0.02$, 0.002, or 0, no correlation emerges. Thus, hard-core potential for both components is required for the $\Delta R-I$ correlation emergency in classic systems, consistent with our understanding.
	
	%chatted********************************	 

	Furthermore, these results suggest that the $\Delta R-I$ correlation may serve as a fingerprint of dark-matter self-interactions. In our model, the correlation appears only when the dark-matter self-interaction cross section exceeds roughly 1\% of the ordinary one. This could provide a novel upper limit on the dark-matter self-interaction cross section from galaxy surveys.
	
	%above is chatted.
	
	%\section{summary}
	
	In summary, using random-interaction ensembles, we demonstrate and explain that the $\Delta R-I$ correlation in low-energy two-component fermionic and classical hard-core many-body systems is not a trivial consequence of many-body symmetries but stems from the short-range attraction of the central force, due to the coordinate transferability of the short-range HO-potential approximation and the virial theorem. In nuclear system, the robust appearance of the $\Delta R-I$ linearity across diverse theoretical frameworks \cite{qvsg-hqxm,MYERS1969395,MYERS1974186,MYERS1980267,PETHICK1996173} serves as a structural fingerprint of the short-range attraction of the nuclear force. Newtonian simulations of bimetallic clusters and galaxies with dark matter support this picture, indicating that the mechanism is neither scale nor quantum specific.
		
	%above is chatted.
	
	While our classical simulations produce the $\Delta R-I$ correlation, experimental verification on bimetallic clusters are still needed. This correlation may also probe dark matter self-interactions with more sophisticated simulations and observational galaxy surveys.
	
	%above is unchatted.
	\begin{acknowledgments}
		The author would like to acknowledge support by China Scholarship Council (Grant No. 202409390020).
	\end{acknowledgments}
	\bibliography{ref}

@article{PhysRevLett.113.171301,
  title = {Mechanism for Thermal Relic Dark Matter of Strongly Interacting Massive Particles},
  author = {Hochberg, Yonit and Kuflik, Eric and Volansky, Tomer and Wacker, Jay G.},
  journal = {Phys. Rev. Lett.},
  volume = {113},
  issue = {17},
  pages = {171301},
  numpages = {5},
  year = {2014},
  month = {Oct},
  publisher = {American Physical Society},
  doi = {10.1103/PhysRevLett.113.171301},
  url = {https://link.aps.org/doi/10.1103/PhysRevLett.113.171301}
}

@article{PhysRevLett.125.131301,
  title = {New Freezeout Mechanism for Strongly Interacting Dark Matter},
  author = {Smirnov, Juri and Beacom, John F.},
  journal = {Phys. Rev. Lett.},
  volume = {125},
  issue = {13},
  pages = {131301},
  numpages = {9},
  year = {2020},
  month = {Sep},
  publisher = {American Physical Society},
  doi = {10.1103/PhysRevLett.125.131301},
  url = {https://link.aps.org/doi/10.1103/PhysRevLett.125.131301}
}

@article{PhysRevLett.128.172001,
  title = {Resonant Self-Interacting Dark Matter from Dark QCD},
  author = {Tsai, Yu-Dai and McGehee, Robert and Murayama, Hitoshi},
  journal = {Phys. Rev. Lett.},
  volume = {128},
  issue = {17},
  pages = {172001},
  numpages = {8},
  year = {2022},
  month = {Apr},
  publisher = {American Physical Society},
  doi = {10.1103/PhysRevLett.128.172001},
  url = {https://link.aps.org/doi/10.1103/PhysRevLett.128.172001}
}

@article{10.1093/mnras/222.2.323,
    author = {Kaiser, Nick},
    title = {Evolution and clustering of rich clusters},
    journal = {Monthly Notices of the Royal Astronomical Society},
    volume = {222},
    number = {2},
    pages = {323-345},
    year = {1986},
    month = {09},
    issn = {0035-8711},
    doi = {10.1093/mnras/222.2.323},
    url = {https://doi.org/10.1093/mnras/222.2.323},
    eprint = {https://academic.oup.com/mnras/article-pdf/222/2/323/18522062/mnras222-0323.pdf},
}

@article{R_F_Casten_1996,
doi = {10.1088/0954-3899/22/11/002},
url = {https://doi.org/10.1088/0954-3899/22/11/002},
year = {1996},
month = {nov},
publisher = {},
volume = {22},
number = {11},
pages = {1521},
author = {R F Casten and N V Zamfir},
title = {The evolution of nuclear structure: the  scheme and related correlations},
journal = {Journal of Physics G: Nuclear and Particle Physics}
}

@Article{Giodini2013,
author={Giodini, S.
and Lovisari, L.
and Pointecouteau, E.
and Ettori, S.
and Reiprich, T. H.
and Hoekstra, H.},
title={Scaling Relations for Galaxy Clusters: Properties and Evolution},
journal={Space Science Reviews},
year={2013},
month={Aug},
day={01},
volume={177},
number={1},
pages={247-282},
issn={1572-9672},
doi={10.1007/s11214-013-9994-5},
url={https://doi.org/10.1007/s11214-013-9994-5}
}

@book{lawson1980theory,
  title={Theory of the Nuclear Shell Model},
  author={Lawson, R.D.},
  isbn={9780198515166},
  lccn={79040480},
  series={Oxford studies in nuclear physics},
  url={https://books.google.co.kr/books?id=dia2AAAAIAAJ},
  year={1980},
  publisher={Clarendon Press}
}

@article{10.1098/rspa.1961.0018,
    author = {Skyrme, T. H. R.},
    title = {A non-linear field theory},
    journal = {Proceedings of the Royal Society of London. A. Mathematical and Physical Sciences},
    volume = {260},
    number = {1300},
    pages = {127-138},
    year = {1961},
    month = {02},
    issn = {0080-4630},
    doi = {10.1098/rspa.1961.0018},
    url = {https://doi.org/10.1098/rspa.1961.0018},
    eprint = {https://royalsocietypublishing.org/rspa/article-pdf/260/1300/127/52613/rspa.1961.0018.pdf},
}

@misc{supp,
    author = {},
    title  = {},
    note   = {See Supplemental Material at URL will be inserted by publisher},
    year   = {}
}

@Article{B703897F,
author ="Tian, Pu",
title  ="Molecular dynamics simulations of nanoparticles",
journal  ="Annu. Rep. Prog. Chem.{,} Sect. C: Phys. Chem.",
year  ="2008",
volume  ="104",
issue  ="0",
pages  ="142-164",
publisher  ="The Royal Society of Chemistry",
doi  ="10.1039/B703897F",
url  ="http://dx.doi.org/10.1039/B703897F"}

@article{BEUTLER1994529,
title = {Avoiding singularities and numerical instabilities in free energy calculations based on molecular simulations},
journal = {Chemical Physics Letters},
volume = {222},
number = {6},
pages = {529-539},
year = {1994},
issn = {0009-2614},
doi = {https://doi.org/10.1016/0009-2614(94)00397-1},
url = {https://www.sciencedirect.com/science/article/pii/0009261494003971},
author = {Thomas C. Beutler and Alan E. Mark and René C. {van Schaik} and Paul R. Gerber and Wilfred F. {van Gunsteren}}
}

@article{Roszkowski_2018,
doi = {10.1088/1361-6633/aab913},
url = {https://doi.org/10.1088/1361-6633/aab913},
year = {2018},
month = {may},
publisher = {IOP Publishing},
volume = {81},
number = {6},
pages = {066201},
author = {Roszkowski, Leszek and Sessolo, Enrico Maria and Trojanowski, Sebastian},
title = {WIMP dark matter candidates and searches-current status and future prospects},
journal = {Reports on Progress in Physics}
}

@article{PhysRevB.54.8398,
  title = {Simple analytical embedded-atom-potential model including a long-range force for fcc metals and their alloys},
  author = {Cai, J. and Ye, Y. Y.},
  journal = {Phys. Rev. B},
  volume = {54},
  issue = {12},
  pages = {8398--8410},
  numpages = {0},
  year = {1996},
  month = {Sep},
  publisher = {American Physical Society},
  doi = {10.1103/PhysRevB.54.8398},
  url = {https://link.aps.org/doi/10.1103/PhysRevB.54.8398}
}

@article{Hale_2018,
doi = {10.1088/1361-651X/aabc05},
url = {https://doi.org/10.1088/1361-651X/aabc05},
year = {2018},
month = {may},
publisher = {IOP Publishing},
volume = {26},
number = {5},
pages = {055003},
author = {Hale, Lucas M and Trautt, Zachary T and Becker, Chandler A},
title = {Evaluating variability with atomistic simulations: the effect of potential and calculation methodology on the modeling of lattice and elastic constants},
journal = {Modelling and Simulation in Materials Science and Engineering}
}

@article{BECKER2013277,
title = {Considerations for choosing and using force fields and interatomic potentials in materials science and engineering},
journal = {Current Opinion in Solid State and Materials Science},
volume = {17},
number = {6},
pages = {277-283},
year = {2013},
note = {Frontiers in Methods for Materials Simulations},
issn = {1359-0286},
doi = {https://doi.org/10.1016/j.cossms.2013.10.001},
url = {https://www.sciencedirect.com/science/article/pii/S1359028613000788},
author = {Chandler A. Becker and Francesca Tavazza and Zachary T. Trautt and Robert A. {Buarque de Macedo}}
}

@misc{nist_interatomic_potentials,  
    author = {Becker, C. A. and Tavazza, F. and Trautt, Z. T. and Buarque~de~Macedo, R. A.},  
    title  = {{Interatomic Potentials Repository}},  
    year   = {2013},  
    url    = {https://www.ctcms.nist.gov/potentials},  
    doi    = {10.18434/m37},  
    note   = {National Institute of Standards and Technology (NIST)},  
    howpublished = {\url{https://www.ctcms.nist.gov/potentials}}  
}

@Article{LAMMPS,
  author = "A. P. Thompson and H. M. Aktulga and R. Berger and 
     D. S. Bolintineanu and W. M. Brown and P. S. Crozier and
     P. J. in 't Veld and A. Kohlmeyer and S. G. Moore and T. D. Nguyen and
     R. Shan and M. J. Stevens and J. Tranchida and C. Trott and S. J. Plimpton",
  title = "{LAMMPS} - a flexible simulation tool for
     particle-based materials modeling at the 
     atomic, meso, and continuum scales",
  journal = "Comp. Phys. Comm.",
  volume =  "271",
  pages =   "108171",
  year =    "2022",
  doi = "10.1016/j.cpc.2021.108171"
}

@article{10.1115/1.3183784,
    author = {Kim, Seong-Gon and Horstemeyer, M. F. and Baskes, M. I. and Rais-Rohani, Masoud and Kim, Sungho and Jelinek, B. and Houze, J. and Moitra, Amitava and Liyanage, Laalitha},
    title = {Semi-Empirical Potential Methods for Atomistic Simulations of Metals and Their Construction Procedures},
    journal = {Journal of Engineering Materials and Technology},
    volume = {131},
    number = {4},
    pages = {041210},
    year = {2009},
    month = {09},
    issn = {0094-4289},
    doi = {10.1115/1.3183784},
    url = {https://doi.org/10.1115/1.3183784},
    eprint = {https://asmedigitalcollection.asme.org/materialstechnology/article-pdf/131/4/041210/5586022/041210\_1.pdf},
}

@article{PhysRevB.70.205411,
  title = {Size-dependent tunneling differential conductance spectra of crystalline Pd nanoparticles},
  author = {Wang, Bing and Wang, Kedong and Lu, Wei and Yang, Jinlong and Hou, J. G.},
  journal = {Phys. Rev. B},
  volume = {70},
  issue = {20},
  pages = {205411},
  numpages = {6},
  year = {2004},
  month = {Nov},
  publisher = {American Physical Society},
  doi = {10.1103/PhysRevB.70.205411},
  url = {https://link.aps.org/doi/10.1103/PhysRevB.70.205411}
}

@article{PhysRevB.75.205107,
  title = {Signatures of random matrix theory in the discrete energy spectra of subnanosize metallic clusters},
  author = {Adams, L. L. A. and Lang, B. W. and Chen, Yu and Goldman, A. M.},
  journal = {Phys. Rev. B},
  volume = {75},
  issue = {20},
  pages = {205107},
  numpages = {11},
  year = {2007},
  month = {May},
  publisher = {American Physical Society},
  doi = {10.1103/PhysRevB.75.205107},
  url = {https://link.aps.org/doi/10.1103/PhysRevB.75.205107}
}

@article{PhysRevLett.77.1123,
  title = {Mesoscopic Fluctuations in the Ground State Energy of Disordered Quantum Dots},
  author = {Sivan, U. and Berkovits, R. and Aloni, Y. and Prus, O. and Auerbach, A. and Ben-Yoseph, G.},
  journal = {Phys. Rev. Lett.},
  volume = {77},
  issue = {6},
  pages = {1123--1126},
  numpages = {0},
  year = {1996},
  month = {Aug},
  publisher = {American Physical Society},
  doi = {10.1103/PhysRevLett.77.1123},
  url = {https://link.aps.org/doi/10.1103/PhysRevLett.77.1123}
}

@article{PhysRevLett.80.4522,
  title = {Statistics of Coulomb Blockade Peak Spacings},
  author = {Patel, S. R. and Cronenwett, S. M. and Stewart, D. R. and Huibers, A. G. and Marcus, C. M. and Duru\"oz, C. I. and Harris, J. S. and Campman, K. and Gossard, A. C.},
  journal = {Phys. Rev. Lett.},
  volume = {80},
  issue = {20},
  pages = {4522--4525},
  numpages = {0},
  year = {1998},
  month = {May},
  publisher = {American Physical Society},
  doi = {10.1103/PhysRevLett.80.4522},
  url = {https://link.aps.org/doi/10.1103/PhysRevLett.80.4522}
}

@article{SKYRME1958615,
title = {The effective nuclear potential},
journal = {Nuclear Physics},
volume = {9},
number = {4},
pages = {615-634},
year = {1958},
issn = {0029-5582},
doi = {https://doi.org/10.1016/0029-5582(58)90345-6},
url = {https://www.sciencedirect.com/science/article/pii/0029558258903456},
author = {T.H.R. Skyrme}
}

@article{10.1098/rspa.1955.0239,
    author = {Elliott, J. P. and Skyrme, T. H. R.},
    title = {Centre-of-mass effects in the nuclear shell-model},
    journal = {Proceedings of the Royal Society of London. A. Mathematical and Physical Sciences},
    volume = {232},
    number = {1191},
    pages = {561-566},
    year = {1955},
    month = {11},
    issn = {0080-4630},
    doi = {10.1098/rspa.1955.0239},
    url = {https://doi.org/10.1098/rspa.1955.0239},
    eprint = {https://royalsocietypublishing.org/rspa/article-pdf/232/1191/561/49334/rspa.1955.0239.pdf},
}

@article{qvsg-hqxm,
  title = {Linear correlations related to neutron skin thickness},
  author = {Lei, Y. and Qi, J. and Lian, X. and Qin, Z. Z. and Bai, C. L.},
  journal = {Phys. Rev. C},
  volume = {113},
  issue = {2},
  pages = {024314},
  numpages = {15},
  year = {2026},
  month = {Feb},
  publisher = {American Physical Society},
  doi = {10.1103/qvsg-hqxm},
  url = {https://link.aps.org/doi/10.1103/qvsg-hqxm}
}

@article{PhysRevC.66.034301,
  title = {Random phase approximation vs exact shell-model correlation energies},
  author = {Stetcu, Ionel and Johnson, Calvin W.},
  journal = {Phys. Rev. C},
  volume = {66},
  issue = {3},
  pages = {034301},
  numpages = {6},
  year = {2002},
  month = {Sep},
  publisher = {American Physical Society},
  doi = {10.1103/PhysRevC.66.034301},
  url = {https://link.aps.org/doi/10.1103/PhysRevC.66.034301}
}

@article{PhysRevC.67.044315,
  title = {Tests of the random phase approximation for transition strengths},
  author = {Stetcu, Ionel and Johnson, Calvin W.},
  journal = {Phys. Rev. C},
  volume = {67},
  issue = {4},
  pages = {044315},
  numpages = {10},
  year = {2003},
  month = {Apr},
  publisher = {American Physical Society},
  doi = {10.1103/PhysRevC.67.044315},
  url = {https://link.aps.org/doi/10.1103/PhysRevC.67.044315}
}

@article{PhysRevC.69.024311,
  title = {Gamow-Teller transitions and deformation in the proton-neutron random phase approximation},
  author = {Stetcu, Ionel and Johnson, Calvin W.},
  journal = {Phys. Rev. C},
  volume = {69},
  issue = {2},
  pages = {024311},
  numpages = {7},
  year = {2004},
  month = {Feb},
  publisher = {American Physical Society},
  doi = {10.1103/PhysRevC.69.024311},
  url = {https://link.aps.org/doi/10.1103/PhysRevC.69.024311}
}

@article{PhysRevC.83.044302,
  title = {Correlations of excited states for $\mathit{sd}$ bosons in the presence of random interactions},
  author = {Lei, Y. and Zhao, Y. M. and Yoshida, N. and Arima, A.},
  journal = {Phys. Rev. C},
  volume = {83},
  issue = {4},
  pages = {044302},
  numpages = {5},
  year = {2011},
  month = {Apr},
  publisher = {American Physical Society},
  doi = {10.1103/PhysRevC.83.044302},
  url = {https://link.aps.org/doi/10.1103/PhysRevC.83.044302}
}

@article{PhysRevC.83.024302,
  title = {Emergence of generalized seniority in low-lying states with random interactions},
  author = {Lei, Y. and Xu, Z. Y. and Zhao, Y. M. and Pittel, S. and Arima, A.},
  journal = {Phys. Rev. C},
  volume = {83},
  issue = {2},
  pages = {024302},
  numpages = {4},
  year = {2011},
  month = {Feb},
  publisher = {American Physical Society},
  doi = {10.1103/PhysRevC.83.024302},
  url = {https://link.aps.org/doi/10.1103/PhysRevC.83.024302}
}

@article{PhysRevC.90.064313,
  title = {Correlations between low-lying yrast states for $sd$ bosons with random interactions},
  author = {Lu, Y. and Zhao, Y. M. and Yoshida, N. and Arima, A.},
  journal = {Phys. Rev. C},
  volume = {90},
  issue = {6},
  pages = {064313},
  numpages = {5},
  year = {2014},
  month = {Dec},
  publisher = {American Physical Society},
  doi = {10.1103/PhysRevC.90.064313},
  url = {https://link.aps.org/doi/10.1103/PhysRevC.90.064313}
}

@article{4qrd-1dqw,
  title = {Novel triaxiality-driven collective feature in atomic nuclei investigated via the two-body random ensemble},
  author = {Fu, G. J. and Qi, Chong},
  journal = {Phys. Rev. C},
  volume = {112},
  issue = {6},
  pages = {L061301},
  numpages = {5},
  year = {2025},
  month = {Dec},
  publisher = {American Physical Society},
  doi = {10.1103/4qrd-1dqw},
  url = {https://link.aps.org/doi/10.1103/4qrd-1dqw}
}

@article{PhysRevResearch.5.013109,
  title = {Structured ground states of randomly interacting bosons},
  author = {White, Charles and Volya, Alexander and Mulhall, Declan and Zelevinsky, Vladimir},
  journal = {Phys. Rev. Res.},
  volume = {5},
  issue = {1},
  pages = {013109},
  numpages = {14},
  year = {2023},
  month = {Feb},
  publisher = {American Physical Society},
  doi = {10.1103/PhysRevResearch.5.013109},
  url = {https://link.aps.org/doi/10.1103/PhysRevResearch.5.013109}
}

@Article{pearson1895,
author={Pearson Karl},
title={VII. Note on regression and inheritance in the case of two parents},
journal={Proc. R. Soc. Lond.},
year={1895},
volume={58},
pages={240-242},
doi={10.1098/rspl.1895.0041},
url={http://doi.org/10.1098/rspl.1895.0041}
}

@Article{Qin2018,
author={Qin, Zhen-Zhen
and Lei, Yang},
title={Predominance of linear Q and $\mu$ systematics in random-interaction ensembles},
journal={Nuclear Science and Techniques},
year={2018},
month={Sep},
day={28},
volume={29},
number={11},
pages={163},
issn={2210-3147},
doi={10.1007/s41365-018-0503-0},
url={https://doi.org/10.1007/s41365-018-0503-0}
}

@article{PhysRevC.93.024319,
  title = {Robust correlations between quadrupole moments of low-lying ${2}^{+}$ states within random-interaction ensembles},
  author = {Lei, Y.},
  journal = {Phys. Rev. C},
  volume = {93},
  issue = {2},
  pages = {024319},
  numpages = {9},
  year = {2016},
  month = {Feb},
  publisher = {American Physical Society},
  doi = {10.1103/PhysRevC.93.024319},
  url = {https://link.aps.org/doi/10.1103/PhysRevC.93.024319}
}

@article{PhysRevC.91.054319,
  title = {Regularities in low-lying states of atomic nuclei with random interactions},
  author = {Fu, G. J. and Shen, J. J. and Zhao, Y. M. and Arima, A.},
  journal = {Phys. Rev. C},
  volume = {91},
  issue = {5},
  pages = {054319},
  numpages = {6},
  year = {2015},
  month = {May},
  publisher = {American Physical Society},
  doi = {10.1103/PhysRevC.91.054319},
  url = {https://link.aps.org/doi/10.1103/PhysRevC.91.054319}
}

@article{PhysRevC.104.054319,
  title = {Robustness of ``noncollective'' rotational behavior for nuclei in the presence of random interactions},
  author = {Shen, J. J. and Jiang, H. and Fu, G. J.},
  journal = {Phys. Rev. C},
  volume = {104},
  issue = {5},
  pages = {054319},
  numpages = {7},
  year = {2021},
  month = {Nov},
  publisher = {American Physical Society},
  doi = {10.1103/PhysRevC.104.054319},
  url = {https://link.aps.org/doi/10.1103/PhysRevC.104.054319}
}

@article{PhysRevLett.84.420,
  title = {Band Structure from Random Interactions},
  author = {Bijker, R. and Frank, A.},
  journal = {Phys. Rev. Lett.},
  volume = {84},
  issue = {3},
  pages = {420--422},
  numpages = {0},
  year = {2000},
  month = {Jan},
  publisher = {American Physical Society},
  doi = {10.1103/PhysRevLett.84.420},
  url = {https://link.aps.org/doi/10.1103/PhysRevLett.84.420}
}

@article{PhysRevC.70.054322,
  title = {Patterns of the ground states in the presence of random interactions: Nucleon systems},
  author = {Zhao, Y. M. and Arima, A. and Shimizu, N. and Ogawa, K. and Yoshinaga, N. and Scholten, O.},
  journal = {Phys. Rev. C},
  volume = {70},
  issue = {5},
  pages = {054322},
  numpages = {8},
  year = {2004},
  month = {Nov},
  publisher = {American Physical Society},
  doi = {10.1103/PhysRevC.70.054322},
  url = {https://link.aps.org/doi/10.1103/PhysRevC.70.054322}
}

@article{PhysRevLett.80.2749,
  title = {Orderly Spectra from Random Interactions},
  author = {Johnson, C. W. and Bertsch, G. F. and Dean, D. J.},
  journal = {Phys. Rev. Lett.},
  volume = {80},
  issue = {13},
  pages = {2749--2753},
  numpages = {0},
  year = {1998},
  month = {Mar},
  publisher = {American Physical Society},
  doi = {10.1103/PhysRevLett.80.2749},
  url = {https://link.aps.org/doi/10.1103/PhysRevLett.80.2749}
}

@article{RevModPhys.81.539,
  title = {Random matrices and chaos in nuclear physics: Nuclear structure},
  author = {Weidenm\"uller, H. A. and Mitchell, G. E.},
  journal = {Rev. Mod. Phys.},
  volume = {81},
  issue = {2},
  pages = {539--589},
  numpages = {0},
  year = {2009},
  month = {May},
  publisher = {American Physical Society},
  doi = {10.1103/RevModPhys.81.539},
  url = {https://link.aps.org/doi/10.1103/RevModPhys.81.539}
}

@article{ZHAO20041,
title = {Regularities of many-body systems interacting by a two-body random ensemble},
journal = {Physics Reports},
volume = {400},
number = {1},
pages = {1-66},
year = {2004},
issn = {0370-1573},
doi = {https://doi.org/10.1016/j.physrep.2004.07.004},
url = {https://www.sciencedirect.com/science/article/pii/S0370157304002972},
author = {Y.M. Zhao and A. Arima and N. Yoshinaga}
}

@article{ZELEVINSKY2004311,
title = {Nuclear structure, random interactions and mesoscopic physics},
journal = {Physics Reports},
volume = {391},
number = {3},
pages = {311-352},
year = {2004},
note = {From atoms to nuclei to quarks and gluons: the omnipresent manybody theory},
issn = {0370-1573},
doi = {https://doi.org/10.1016/j.physrep.2003.10.008},
url = {https://www.sciencedirect.com/science/article/pii/S0370157303004319},
author = {Vladimir Zelevinsky and Alexander Volya}
}

@article{KOTA2001223,
title = {Embedded random matrix ensembles for complexity and chaos in finite interacting particle systems},
journal = {Physics Reports},
volume = {347},
number = {3},
pages = {223-288},
year = {2001},
issn = {0370-1573},
doi = {https://doi.org/10.1016/S0370-1573(00)00113-7},
url = {https://www.sciencedirect.com/science/article/pii/S0370157300001137},
author = {V.K.B. Kota}
}

@article{PETHICK1996173,
title = {The dependence of neutron skin thickness and surface tension on neutron excess},
journal = {Nuclear Physics A},
volume = {606},
number = {1},
pages = {173-182},
year = {1996},
issn = {0375-9474},
doi = {https://doi.org/10.1016/0375-9474(96)00216-3},
url = {https://www.sciencedirect.com/science/article/pii/0375947496002163},
author = {C.J. Pethick and D.G. Ravenhall}
}

@article{MYERS1980267,
title = {Droplet-model theory of the neutron skin},
journal = {Nuclear Physics A},
volume = {336},
number = {2},
pages = {267-278},
year = {1980},
issn = {0375-9474},
doi = {https://doi.org/10.1016/0375-9474(80)90623-5},
url = {https://www.sciencedirect.com/science/article/pii/0375947480906235},
author = {W.D. Myers and W.J. Swiatecki}
}

@article{MYERS1974186,
title = {The nuclear droplet model for arbitrary shapes},
journal = {Annals of Physics},
volume = {84},
number = {1},
pages = {186-210},
year = {1974},
issn = {0003-4916},
doi = {https://doi.org/10.1016/0003-4916(74)90299-1},
url = {https://www.sciencedirect.com/science/article/pii/0003491674902991},
author = {W.D Myers and W.J Swiatecki}
}

@article{MYERS1969395,
title = {Average nuclear properties},
journal = {Annals of Physics},
volume = {55},
number = {3},
pages = {395-505},
year = {1969},
issn = {0003-4916},
doi = {https://doi.org/10.1016/0003-4916(69)90202-4},
url = {https://www.sciencedirect.com/science/article/pii/0003491669902024},
author = {William D Myers and W.J Swiatecki}
}

@article{PhysRevLett.131.202302,
  title = {Determination of the Neutron Skin of $^{208}\mathrm{Pb}$ from Ultrarelativistic Nuclear Collisions},
  author = {Giacalone, Giuliano and Nijs, Govert and van der Schee, Wilke},
  journal = {Phys. Rev. Lett.},
  volume = {131},
  issue = {20},
  pages = {202302},
  numpages = {6},
  year = {2023},
  month = {Nov},
  publisher = {American Physical Society},
  doi = {10.1103/PhysRevLett.131.202302},
  url = {https://link.aps.org/doi/10.1103/PhysRevLett.131.202302}
}

@article{PhysRevLett.129.042501,
  title = {Precision Determination of the Neutral Weak Form Factor of $^{48}\mathrm{Ca}$},
  author = {Adhikari, D. and Albataineh, H. and Androic, D. and Aniol, K. A. and Armstrong, D. S. and Averett, T. and Ayerbe Gayoso, C. and Barcus, S. K. and Bellini, V. and Beminiwattha, R. S. and Benesch, J. F. and Bhatt, H. and Bhatta Pathak, D. and Bhetuwal, D. and Blaikie, B. and Boyd, J. and Campagna, Q. and Camsonne, A. and Cates, G. D. and Chen, Y. and Clarke, C. and Cornejo, J. C. and Covrig Dusa, S. and Dalton, M. M. and Datta, P. and Deshpande, A. and Dutta, D. and Feldman, C. and Fuchey, E. and Gal, C. and Gaskell, D. and Gautam, T. and Gericke, M. and Ghosh, C. and Halilovic, I. and Hansen, J.-O. and Hassan, O. and Hauenstein, F. and Henry, W. and Horowitz, C. J. and Jantzi, C. and Jian, S. and Johnston, S. and Jones, D. C. and Kakkar, S. and Katugampola, S. and Keppel, C. and King, P. M. and King, D. E. and Kumar, K. S. and Kutz, T. and Lashley-Colthirst, N. and Leverick, G. and Liu, H. and Liyanage, N. and Mammei, J. and Mammei, R. and McCaughan, M. and McNulty, D. and Meekins, D. and Metts, C. and Michaels, R. and Mihovilovic, M. and Mondal, M. M. and Napolitano, J. and Narayan, A. and Nikolaev, D. and Owen, V. and Palatchi, C. and Pan, J. and Pandey, B. and Park, S. and Paschke, K. D. and Petrusky, M. and Pitt, M. L. and Premathilake, S. and Quinn, B. and Radloff, R. and Rahman, S. and Rashad, M. N. H. and Rathnayake, A. and Reed, B. T. and Reimer, P. E. and Richards, R. and Riordan, S. and Roblin, Y. R. and Seeds, S. and Shahinyan, A. and Souder, P. and Thiel, M. and Tian, Y. and Urciuoli, G. M. and Wertz, E. W. and Wojtsekhowski, B. and Yale, B. and Ye, T. and Yoon, A. and Xiong, W. and Zec, A. and Zhang, W. and Zhang, J. and Zheng, X.},
  collaboration = {CREX Collaboration},
  journal = {Phys. Rev. Lett.},
  volume = {129},
  issue = {4},
  pages = {042501},
  numpages = {8},
  year = {2022},
  month = {Jul},
  publisher = {American Physical Society},
  doi = {10.1103/PhysRevLett.129.042501},
  url = {https://link.aps.org/doi/10.1103/PhysRevLett.129.042501}
}

@article{PhysRevC.46.1825,
  title = {Neutron radii of the calcium isotopes},
  author = {Gibbs, W. R. and Dedonder, J.-P.},
  journal = {Phys. Rev. C},
  volume = {46},
  issue = {5},
  pages = {1825--1833},
  numpages = {0},
  year = {1992},
  month = {Nov},
  publisher = {American Physical Society},
  doi = {10.1103/PhysRevC.46.1825},
  url = {https://link.aps.org/doi/10.1103/PhysRevC.46.1825}
}

@article{ refId0,
	author = {{Rein, H.} and {Liu, S.-F.}},
	title = {REBOUND: an open-source multi-purpose $N$-body code   for collisional dynamics},
	DOI= "10.1051/0004-6361/201118085",
	url= "https://doi.org/10.1051/0004-6361/201118085",
	journal = {A{\&}A},
	year = 2012,
	volume = 537,
	pages = "A128",
	month = "",
}
	
\end{document}